\documentclass[prl,twocolumn,showpacs,preprintnumbers,amsmath,amssymb]{revtex4-1}
\usepackage[latin1]{inputenc}
\usepackage[T1]{fontenc}

\usepackage{graphicx}

\newcommand{\XeqY}[2]{$#1\!\!=\!\!#2$}

\begin{document}

\preprint{APS/Precision Mass Measurements beyond $^{132}$Sn}
\title{Precision Mass Measurements beyond $^{132}$Sn: 
  Anomalous behaviour of odd-even staggering of binding energies}

\author{J.~Hakala}\email{jani.hakala@phys.jyu.fi}
\author{J.~Dobaczewski}
\author{D.~Gorelov}
\author{T.~Eronen} \thanks{\emph{Present address:} Max-Planck-Institut f\"{u}r Kernphysik, Saupfercheckweg 1, 69117 Heidelberg, Germany} 
\author{A.~Jokinen}
\author{A.~Kankainen}
\author{V.S.~Kolhinen}		
\author{M.~Kortelainen}
\author{I.~D.~Moore}
\author{H.~Penttil\"{a}}
\author{S.~Rinta-Antila}
\author{J.~Rissanen}
\author{A.~Saastamoinen}
\author{V.~Sonnenschein}
\author{J.~\"{A}yst\"{o}}\email{juha.aysto@phys.jyu.fi}
\affiliation{Department of Physics, P.O. Box 35 (YFL), FI-40014 
  University of Jyv\"{a}skyl\"{a}, Finland}

\date{\today}

\begin{abstract}

Atomic masses of the neutron-rich isotopes $^{121-128}$Cd, 
$^{129,131}$In, $^{130-135}$Sn, $^{131-136}$Sb, and $^{132-140}$Te have
been measured with high precision (10 ppb) using the Penning trap mass
spectrometer JYFLTRAP. Among these, the masses of four r-process
nuclei $^{135}$Sn, $^{136}$Sb, and $^{139,140}$Te were measured for
the first time. An empirical neutron pairing gap expressed as the
odd-even staggering  of isotopic masses shows a strong quenching
across \XeqY{N}{82} for Sn, with the $Z$-dependence that is
unexplainable by the current theoretical models. 

\end{abstract}

 	%% PACS, the Physics and Astronomy
			  		%% Classification Scheme.
\maketitle

The doubly magic $^{132}$Sn nucleus has been probed intensively by
nuclear spectroscopy over the last two decades. It has been found to
exhibit features of exceptional purity for its single particle
structure \cite{Hoff1996,Jones2010}. This provides an ideal
starting point for exploring detailed evolution of nuclear
structure of more neutron-rich nuclei beyond the \XeqY{N}{82} closed
shell in the vicinity of Sn.  Only a few experimental and theoretical
attempts along these lines have been performed recently. No experimental 
data exist for excited states or masses for nuclides below Sn with $N>82$. 
The experimental situation is slightly better for $Z>50$ isotopes of
Sb and Te because of their easier access.

Recent data on the B(E2) transition strengths for $^{132}$Te,
$^{134}$Te and $^{136}$Te isotopes \cite{Radford2002} and their
interpretation using a quadrupole-plus-pairing Hamiltonian and
Quasiparticle Random Phase Approximation (QRPA) \cite{Terasaki2002}
suggested the need for reduced neutron pairing to explain the
observed anomalous asymmetry in the B(E2) values across the
\XeqY{N}{82} neutron shell. This behaviour was not observed in
standard shell model calculations \cite{Radford2002}.
Another shell model calculation of the binding energies of heavy Sn
isotopes with $A>133$ \cite{Kartamyshev2007} suggested the importance of
pairing correlations and the strength of the pairing interaction in
general for weakly bound nuclei. Therefore, it would be necessary
to probe the evolution of odd-even staggering of masses
\cite{Satula1998} around the \XeqY{N}{82} neutron shell to learn about
the magnitude of pairing and its variation as a function of $Z$ and $N$
beyond $^{132}$Sn. 

\begin{figure}[htbp]
  \centering
  \includegraphics[width=0.99\columnwidth]{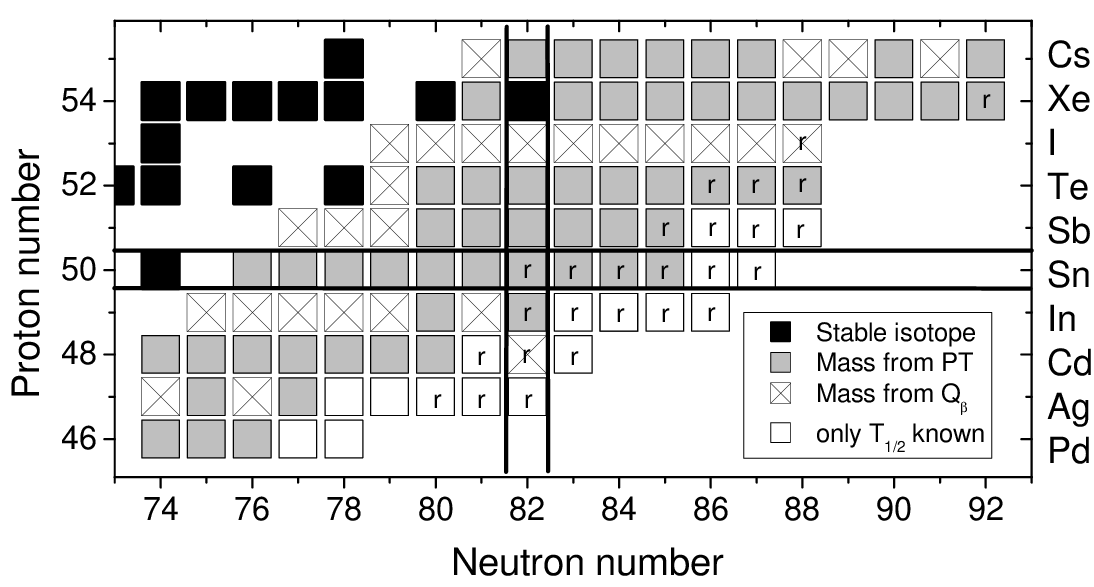}
  \caption{\label{fig:area} Neutron-rich isotopes with $T_z\!\ge\!13$
    whose masses have been determined by Penning trap (PT) or
    $Q_{\beta}$-measurements. Letter
    \textit{r} denotes r-process nuclei according to 
    Ref.~\cite{Arnould2007}.}
\end{figure}

High precision of present-day ion-trap mass spectrometry combined with high  
sensitivity \cite{Blaum2006} can provide the needed information on
mass differences such as  one- and two-nucleon separation energies,
shell gaps and empirical pairing energies. For example, the masses of
neutron-rich Sn and Xe isotopes were recently measured up to
$^{134}$Sn and $^{146}$Xe with a Penning trap mass spectrometer
ISOLTRAP at the CERN ISOLDE facility \cite{Dworschak2008,Neidherr2009}.
In this Letter we wish to
present new data of high-precision mass measurements of neutron-rich
Cd, In, Sn, Sb, and Te isotopes across the \XeqY{N}{82} neutron shell
by using the JYFLTRAP Penning trap. These nuclides are also of
interest for nuclear astrophysics models of element synthesis,  in
particular, to explain the large r-process abundance peak at
\XeqY{A}{130} \cite{Arnould2007}, see Fig.~\ref{fig:area}. 
In more general context, a vast body of nuclear data on neutron-rich
isotopes is needed for r-process nucleosynthesis predictions. Such
data include masses, single particle spectra, pairing characteristics
as well as decay properties and reaction rates. In all of these the
binding energies or masses of ground, isomeric and excited states play
key roles \cite{Arnould2007}.
 
The measurements were performed using the JYFLTRAP Penning trap
mass spectrometer \cite{Jokinen2006} which is connected
to the Ion Guide Isotope Separator On-Line (IGISOL) mass separator
\cite{Penttila2005}. The ions of interest were produced in
proton-induced fission reactions by bombarding a natural uranium
target with a proton beam of 25 MeV energy.  
A thorium target was used in the case of $^{129}$In and isotopes of
Sb. 

Fission products stopped in a helium-filled gas cell at a pressure of
about 200 mbar as singly-charged ions were transported out of the gas
cell, accelerated to 30 keV energy, and mass separated. A gas-filled
radio frequency quadrupole cooler and buncher prepared the ions for
the Penning trap setup. 

In a Penning trap an ion has three different eigenmotions: axial
motion $(\nu_{z})$ and two radial motions, magnetron $(\nu_{-})$ and
modified cyclotron $(\nu_{+})$ motion. The frequencies of the radial
motions sum in first order to the cyclotron frequency
$\nu_c\!=\!\frac{1}{2\pi}\frac{q}{m}B$. Here \emph{q} and \emph{m} are
the charge and mass of the ion, and \emph{B} is the magnetic field.  

JYFLTRAP consists of two cylindrical Penning traps, the purification
and precision trap, which are located inside a 7-T superconducting magnet. 
Both traps were used to purify the samples, first with the sideband
cooling technique \cite{Savard1991} for isobaric separation and then,
if prompted by the presence of isomers or other contaminants, with the 
Ramsey cleaning technique \cite{Eronen2008a}. After a cleaned
sample was trapped in the precision trap, the time-of-flight
ion-cyclotron resonance (TOF-ICR) technique \cite{Konig1995} was
applied in order to determine the resonance frequency. The Ramsey
method of separated oscillatory fields \cite{George2007a}, which makes
the sidebands more pronounced and narrower, was used in the majority
of measurements. The experimental setup and the measurement technique are
described in more detail in Ref.~\cite{Eronen2012}.

\begin{figure}[htbp]
  \centering
  \includegraphics[width=0.99\columnwidth]{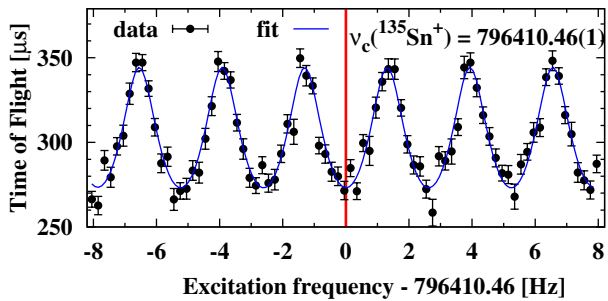}
  \caption{\label{fig:sn135} (Color online) A time-of-flight (TOF) resonance
    from the JYFLTRAP setup for $^{135}\mathrm{Sn}^{\!+}$. A
    two-fringe Ramsey pattern of 25-350-25 ms (\textit{on-off-on}) was
    used in this case.}
\end{figure}

A sample resonance is shown in Fig.~\ref{fig:sn135}. 
When the cyclotron frequencies of an unknown species
($m_{\mathrm{meas}}$) and a well-known reference ion
($m_{\mathrm{ref}}$) are known, and both are singly 
charged ions, the atomic mass can be determined:
\begin{equation}
  m_{\mathrm{meas}} = \frac{\nu_{c,\mathrm{ref}}}{\nu_{c,\mathrm{meas}}}
  \left(m_{\mathrm{ref}} - m_{e}\right) + m_e,
\end{equation}
where $m_{e}$ is the mass of the electron.  The data analysis
procedure was almost identical compared to \cite{Hakala2011}. In this
work, the systematic uncertainties due to the magnetic field
fluctuations were minimised by applying so-called interleaved
frequency scanning described in Ref.~\cite{Eronen2009a}.

The results are given in Table \ref{tab:results}. Excitation times
between 100 ms and 800 ms were used depending on the half-life of the
isotope. In the case of Ramsey excitations, excitation patterns with two
25 ms pulses separated by a waiting time, were used. Table
\ref{tab:results} lists only the masses of the ground states which are
relevant for the further discussion of the results. A paper containing
the isomeric data will be submitted separately. The new data agrees
with the earlier ion-trap measurements and presents significant
improvement in accuracy for all Sn, Sb and Te  isotopes beyond
\XeqY{N}{82}.

\begin{table}[htbp]
  \caption{\label{tab:results}
    Cyclotron frequency ratios $\bar{r}$ and ground state mass excess
    values in keV based on this work. Masses of reference Xe isotopes
    in column 1 for given $A$ are from     
    \cite{Audi2003,Redshaw2009,Redshaw2009b}.
    Results from other direct mass measurements are from
    ISOLTRAP \cite{Sikler2005,Dworschak2008,Breitenfeldt2010}
    or an experimental storage ring (ESR) at the fragment separator
    (FRS) \cite{Sun2008}. Otherwise, the value from Atomic
    Mass Evaluation 2003 (AME2003) \cite{Audi2003}
    is given.}
  \begin{ruledtabular}
    \begin{tabular}{cllllc}
      Xe & nuclide
      & $\bar{r}=\frac{\nu_{c,\mathrm{ref}}}{\nu_{c,\mathrm{meas}}}$ & 
      JYFLTRAP & literature & ref.\\
      $A$ & & & keV & keV \\[2pt]
      \hline
      && \\[-6pt]
130 & $^{121}$Cd$^{\dagger}$ & 0.930790292(23) & -81074.2(28) & -81060(80) 
 & \cite{Audi2003} \\ 
" & $^{122}$Cd & 0.938492175(22) & -80610.8(27) & -80616.6(44) 
 & \cite{Breitenfeldt2010} \\
" & $^{123}$Cd$^{\dagger}$  & 0.946216645(22) & -77414.4(26) & -77367(93) 
 & \cite{Breitenfeldt2010} \\
" & $^{124}$Cd & 0.953920584(26) & -76702(4) & -76697(10) 
 & \cite{Breitenfeldt2010} \\
" & $^{125}$Cd$^{\dagger}$  & 0.961646357(24) & -73348.1(29) & -73360(70) 
 & \cite{Audi2003} \\
" & $^{126}$Cd & 0.969353430(25) & -72257(3) & -72256.5(42)
 & \cite{Breitenfeldt2010} \\
" & $^{127}$Cd & 0.97708259(11) & -68493(13) & -68520(70) 
 & \cite{Audi2003} \\
" & $^{128}$Cd & 0.984791049(83) & -67234(10) & -67250(17)
 & \cite{Breitenfeldt2010} \\
" & $^{129}$In$^{\dagger}$ & 0.992442788(22) & -72838.0(26) & -72940(40) 
 & \cite{Audi2003} \\
" & $^{131}$In$^{\dagger}$ & 1.007878672(22) & -68025.0(26) & -68137(28) 
 & \cite{Audi2003} \\
" & $^{130}$Sn$^{\dagger}$  & 1.000080552(28) & -80133(4) & -80134(16) 
 & \cite{Sikler2005} \\
132 & $^{131}$Sn & 0.992516774(26) & -77262(20)$^{\mathrm{x}}$ & -77264(10) 
 & \cite{Dworschak2008} \\
" & $^{132}$Sn & 1.000103654(26) & -76543(4) & -76547(7) 
  & \cite{Dworschak2008} \\
134 & $^{133}$Sn & 0.992670308(18) & -70874.4(24) & -70847(23) 
  & \cite{Dworschak2008} \\
" & $^{134}$Sn & 1.000173910(25) & -66432(4) & -66320(150) 
 & \cite{Dworschak2008} \\
130 & $^{135}$Sn & 1.038731983(25) & -60632(3) & \\
" & $^{131}$Sb & 1.007763324(18) & -81982.5(21) & -81988(21) & \cite{Audi2003} \\
" & $^{132}$Sb$^{\dagger}$ & 1.015480774(22) & -79635.6(27) & -79674(14) & \cite{Audi2003} \\ 
" & $^{133}$Sb & 1.023184733(31) & -78921(4) & -78943(25) & \cite{Audi2003} \\ 
" & $^{134}$Sb$^{\dagger}$ & 1.030923280(17) & -74021.1(21) & -74170(40) & \cite{Audi2003} \\ 
" & $^{135}$Sb & 1.038657131(24) & -69689.6(29) & -69710(100) & \cite{Audi2003} \\ 
" & $^{136}$Sb & 1.046397992(52) & -64510(7) & \\
" & $^{132}$Te & 1.015434872(33) & -85190(4) & -85182(7) & \cite{Audi2003} \\ 
" & $^{133}$Te$^{\dagger}$ & 1.023151534(18) & -82938.2(22) & -82945(24) & \cite{Audi2003} \\ 
" & $^{134}$Te & 1.030852922(27) & -82535(4) & -82559(11) & \cite{Audi2003} \\ 
" & $^{135}$Te & 1.038590701(22) & -77727.9(26) & -77725(123) & \cite{Audi2003} \\ 
" & $^{136}$Te & 1.046316045(24) & -74425.7(29) & -74430(50) & \cite{Audi2003} \\ 
" & $^{137}$Te & 1.054056424(21) & -69304.2(25) & -69290(120) & \cite{Sun2008} \\ 
" & $^{138}$Te & 1.061784295(36) & -65696(5) & -65755(122) 
 & \cite{Sun2008} \\
" & $^{139}$Te & 1.069527729(29) & -60205(4) & \\
" & $^{140}$Te & 1.07725759(23) & -56357(27) & \\
    \end{tabular}
    \begin{flushleft}
      x corrected for 65.1 keV isomer, see Ref.~\cite{Audi2003} \\
      $\dagger$ isomer separated
    \end{flushleft}
  \end{ruledtabular}
\end{table}
\clearpage
%
% \section{Discussion}
% \label{sec:discussion}

\begin{figure}[htbp]
  \centering
  \includegraphics[width=0.99\columnwidth]{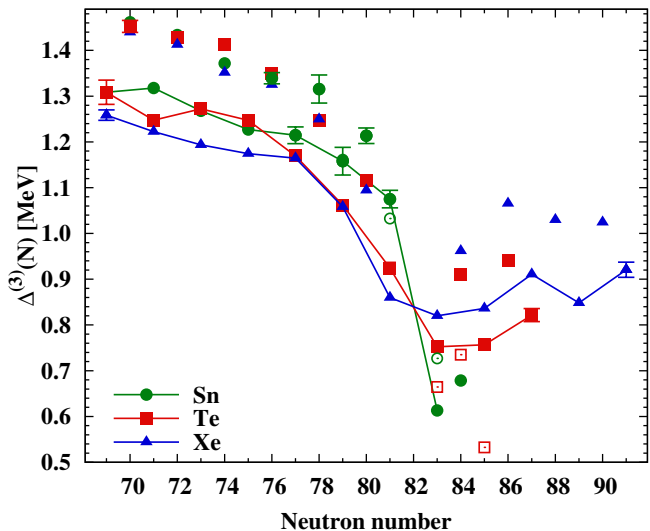}
  \caption{\label{fig:expstaggering} (Color online) Experimental
    odd-even mass staggering for Sn, Te, and Xe isotope chains.
    For clarity, odd-$N$ points have been connected with lines and 
    the error bars have been omitted when $\delta(\Delta^{(3)}) < 10$
    keV. The open symbols present the experimental values around the
    shell-gap prior to this work. The data points for \XeqY{N}{82} are
    beyond the scale.}  
\end{figure}

The high precision of Penning-trap mass measurements enables for the
first time a critical evaluation of odd-even staggering of binding
energies and related empirical pairing gaps across the \XeqY{N}{82}
shell gap.
The most simple example is the three-point odd-even-mass-staggering
formula \cite{Satula1998} 
\begin{equation}
  \Delta^{(3)}(N) = (-1)^{N} [E(N+1) - 2 E(N) + E(N-1)]/2,
\end{equation}
where $N$ is the number of nucleons (neutrons or protons) and $E$ the
binding energy. The $\Delta^{(3)}$ staggering mostly depends on the
intensity of pairing correlations in nuclei, but, as we discuss below,
it is also affected by the polarisation effects.

Figure \ref{fig:expstaggering} shows the experimental neutron $\Delta^{(3)}$
staggering for Sn, Te, and Xe isotopes crossing the \XeqY{N}{82} shell
closure. The difference between the values at \XeqY{N}{81} and 83
shows a large asymmetry for Sn but a much smaller one for Te and
Xe. This indicates a considerably stronger quenching in pairing gap
for Sn than for Te and Xe suggesting the importance of core
polarisation effects. A similar asymmetry observed for B(E2)
values of n-rich Te isotopes was also traced to reduced neutron
pairing above the \XeqY{N}{82} shell closure \cite{Terasaki2002}.

In order to probe this question theoretically, we performed
self-consistent calculations of varying complexity, by using the Sly4
\cite{Chabanat1998} 
energy density functional and contact pairing force. To make a
meaningful comparison, in each variant of the calculation, the pairing
strength (equal for neutrons and protons) was adjusted so as to give
the average pairing gap in $^{120}$Sn equal to 1.245 MeV. The pairing
channel was described within the Hartree-Fock-Bogoliubov (HFB)
approximation and the blocking and filling approximations
\cite{Bertsch2009,Schunck2010} were used to treat odd nuclei. 

\begin{figure}[htpb]
  \centering
  \includegraphics[width=0.99\columnwidth]{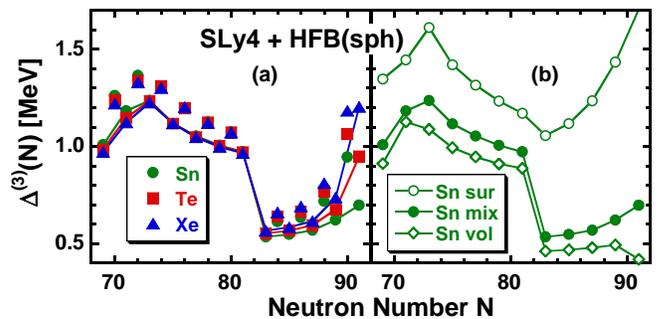}
  \caption{\label{fig:staggering1} (Color online) Same as in
    Fig.~\ref{fig:expstaggering} but for the calculated (spherical HFB)
    odd-even mass staggering in (a) Sn, Te, and Xe isotope chains and
    mixed pairing force, and (b) Sn isotope chain and surface, mixed,
    and volume pairing forces.}
\end{figure}

First, to provide a baseline for further analyses, in 
Fig.~\ref{fig:staggering1} 
we show the neutron $\Delta^{(3)}$ staggering calculated within the
spherical approximation \cite{Dobaczewski1984}. 
Such an approximation allows us to look at pure effects of pairing
correlations, whereby the binding energies of odd isotopes are
decreased due to one pair being broken. Otherwise, in this
approximation, effects due to deformation polarisations exerted by
valence particles are completely switched off.
From Fig.~\ref{fig:staggering1}b we see that for the volume (vol) or
mixed (mix) pairing forces \cite{Dobaczewski2002}, the experimental
decrease of $\Delta^{(3)}$ when crossing the \XeqY{N}{82} gap in Sn is
very well reproduced, whereas the data exclude the pure surface
pairing force. 
Such pairing decrease is due to a lower level density above the
\XeqY{N}{82} gap \cite{Terasaki2002} (the surface pairing force is
unable to discriminate between the two surface-type orbitals
$\nu$h$_{11/2}$ and $\nu$f$_{7/2}$ located below and above the
\XeqY{N}{82} shell gap, respectively). Spherical calculations miss the
experimental values in Te and Xe isotopes
(Fig.~\ref{fig:staggering1}a), which indicates that the polarisation
effects must here be taken into account explicitly.

\begin{figure}[htbp]
  \centering
  \includegraphics[width=0.99\columnwidth]{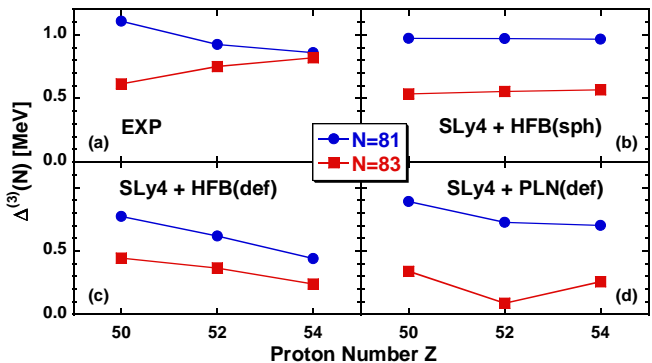}
  \caption{\label{fig:staggering2} (Color online)
    Odd-even staggering for the \XeqY{N}{81} and 83 isotones
    as measured in experiment (a) and estimated within the spherical (b),
    deformed (c), and deformed particle-number-conserving (d) self-consistent
    calculations.}
\end{figure}

To better illustrate the trends in $Z>50$ nuclei, in
Fig.~\ref{fig:staggering2} we show the experimental neutron
$\Delta^{(3)}$ staggering in Sn, Te, and Xe isotopes along 
the \XeqY{N}{81} and 83 lines, compared with theoretical results. In
experiment (Fig.~\ref{fig:staggering2}a), the difference between the
\XeqY{N}{81} and 83 isotones 
smoothly decreases from about 0.5 MeV in Sn to almost zero in Xe,
whereas spherical results (discussed above and repeated in
Fig.~\ref{fig:staggering2}b) show no such a decrease at all. 

To analyse the effects of polarisations induced by deformation,
we have performed two additional HFB calculations. First, by
using the code HFODD (v2.51i) \cite{Schunck2012}, we allowed valence
particles or holes to induce self-consistently deformed
shapes. Then, the even isotones, \XeqY{N}{80}, 82, and 84, turn out
to be spherical anyhow, namely, neither the closed-core \XeqY{N}{82}
systems become deformed, nor the paired two neutron particles
or holes induce non-zero deformations. Only the polarisation
effects exerted by unpaired odd neutron particles or holes are
strong enough to induces non-zero deformations in the odd
isotones, \XeqY{N}{81} and 83.

In this way, the odd-particle polarisations lead to lower
values of the $\Delta^{(3)}$ staggering solely through
increased binding energies of odd isotones. Since such
polarisation increases with adding protons, the obtained
values of $\Delta^{(3)}$ decrease with $Z$, as illustrated in
Fig.~\ref{fig:staggering2}c. However, this trend does not
depend on whether deformation is induced by odd particles or
holes; therefore, on both sides of the \XeqY{N}{82} shell gap we
obtain identical decrease with $Z$, at variance with experiment.

To test if the above results may be affected by the particle-number
non-conservation, inherent to the HFB theory, we have repeated all
calculations by using the HFBTHO code
\cite{Stoitsov2005,Stoitsov2007}. Here, within the Lipkin-Nogami
method, we were able to include the approximate particle-number
projection after variation. Moreover, the obtained solutions were next
exactly projected on good particle numbers. Such a projected
Lipkin-Nogami (PLN) method was extensively tested \cite{Stoitsov2007}
and proved to be very efficient in describing pairing correlations
in near-closed-shell systems.

The obtained results are shown in Fig.~\ref{fig:staggering2}d.
We see that the general pattern is not qualitatively changed.
The dynamic pairing correlations, which are now nonzero even
in the \XeqY{N}{82} isotones, lead to larger values of the
$\Delta^{(3)}$ staggering, which in the \XeqY{N}{81} isotones
perfectly well reproduce the experimental trend. However, in
the \XeqY{N}{83} isotones, the disagreement with data remains a
puzzle. Indeed, the asymmetry of the trend of the staggering,
measured below and above the \XeqY{N}{82} gap, points to specific
effects related to orbitals occupied beyond \XeqY{N}{82} or to the
their weak binding, which are not captured by the current
state-of-the-art theoretical approaches.

\begin{acknowledgments}
  This work has been supported by the Academy of Finland under the
  Centre of Excellence Programme 2006--2014 (Nuclear and
  Accelerator Based Physics Programme at JYFL) and the FIDIPRO program.
\end{acknowledgments}

%%\bibliographystyle{natbib}
%%\bibliography{fission}
%%\input{final.bbl}
%%\input{article.bbl}

%Merlin.mbs v4.21 2009-07-09.
%

\end{document}